\documentclass[usegraphicx,usenatbib]{mn2e}

\usepackage{fixltx2e}
\usepackage{mathptmx}

\include{defn}

\def\ApJ{{\em Astrophys.~J.}}

\def\MN{{\em Mon.~Not.~R.~astr.~Soc.}}
\def\Nat{{\em Nature}}

\def\etal{{et al.\thinspace}}

\def\spose#1{\hbox to 0pt{#1\hss}}

\def\multleft#1{\hbox to size{\vbox {\halign {\lft{##}\cr #1}}\hfill}\par}
\def\multright#1{\hbox to size{\vbox {\halign {\rt{##}\cr #1}}\hfill}\par}

\def\boxit#1{\vbox{\hrule\hbox{\vrule\kern3pt\vbox{\kern3pt
          #1 \kern3pt}\kern3pt\vrule}\hrule}}

\def\cm{{\rm\thinspace cm}}

\def\erg{{\rm\thinspace erg}}

\def\ga{{\rm\thinspace gauss}}

\def\keV{{\rm\thinspace keV}}

\def\Msun{\hbox{$\rm\thinspace M_{\odot}$}}

\def\s{{\rm\thinspace s}}

\def\pcmsq{\hbox{$\cm^{-2}\,$}}

\def\ps{\hbox{$\s^{-1}\,$}}

\vspace{-2cm}

\begin{document}

\title{Constraints on the absorption-dominated model for the X-ray spectrum of MCG--6-30-15}
\author[C.S. Reynolds et al. ]{
C.~S.~Reynolds$^1$\thanks{E-mail: chris@astro.umd.edu},
A.~C.~Fabian$^2$, L.~W.~Brenneman$^3$, G.~Miniutti$^4$, P.~Uttley$^5$,
and L.~C.~Gallo$^6$ \\
  $^1$Department of Astronomy and the Maryland Astronomy Center for 
Theory and Computation, University of Maryland, College Park, MD~20742, USA\\
  $^2$Institute of Astronomy, Madingley Road, Cambridge. CB3 0HA\\
  $^3$NASA/Goddard Space Flight Center, Greenbelt, MD, USA\\
  $^4$LAEX, Centro de Astrobiologia (CSIC--INTA); LAEFF, P.O: Box 78, E-28691, Villanueva de la Ca\~nada, Madrid, Spain\\
  $^5$School of Physics and Astronomy, University of Southampton, Southampton, SO17~1BJ\\
  $^6$Dept. of Astronomy and Physics, Saint Mary's University, 923 Robie 
  Street, Halifax, NS, B3H 3C3, Canada}
\maketitle
  
\begin{abstract}
Complexities in the X-ray spectrum of the nearby Seyfert 1.2 galaxy
MCG--6-30-15 are commonly interpreted in terms of a broad iron line
and the associated Compton reflection hump from the innermost
relativistic regions of an accretion disk around a rapidly spinning
black hole.  However, an alternative model has recently been
proposed in which these spectral features are caused entirely by
complex (ionized and partial-covering) absorption.  By considering the
fluorescent emission that must accompany photoelectric absorption, we
show that the absorption-dominated model over-predicts the 6.4\,keV iron
line flux unless the marginally Compton-thick absorber responsible for
the hard X-ray hump satisfies very restrictive geometric constraints.
In the absence of a specific model that both obeys these geometrical
constraints and is physically-plausible, the relativistic-reflection
model is favoured.
\end{abstract}


\section{Introduction}

The X-ray spectrum of the nearby Seyfert 1.2 galaxy MCG--6-30-15
displays the archetypal relativistic broad iron line (Tanaka et
al. 1995).  The line shape is well explained by relativistic blurring
(Fabian et al 1989) of the X-ray reflection spectrum expected from the
inner accretion disc around the central black hole. The red wing of
the line is so broad that extreme redshifts are required indicating
that the inner radius of the disc is close to the black hole and thus
that the hole is rapidly spinning (Dabrowski et al 1997; Brenneman \&
Reynolds 2006). In this case, much of the emission originates from the
region where gravity is so strong that light bending is a large
effect. This in turn can explain the apparently disconnected spectral
behaviour of the source (Miniutti et al 2003; Miniutti \& Fabian
2004), with the relativistic-reflection component showing much less
variation compared with the power-law continuum (Matsumoto et al 2003;
Shih et al 2002; Fabian \& Vaughan 2003; Larsson et al 2007).  A key
prediction of the disk reflection model is the presence of a powerful
Compton reflection hump at $\sim 20\keV$ which varies in step with the
broad iron line; this is indeed seen in the {\it Suzaku} data for this
object (Miniutti et al. 2007).

Miller, Turner \& Reeves (2008, hereafter MTR) propose an alternative
explanation in which the X-ray spectrum of MCG--6-30-15 is sculpted by
multiple absorbers.  A moderately photoionized absorber
partially-covers the primary power-law continuum source producing a
3--7\,keV hump in the spectrum which mimics the red-wing of a
broadened iron line.  The weak narrow iron line present in the
spectrum is described by reflection from distant material, but that
reflection spectrum must then itself be absorbed by a second
photoionized and marginally-Compton thick structure in order to
correctly model the powerful 20\,keV hump seen in the {\it Suzaku}
spectrum.  The MTR model requires no relativistic effects and has most
zones at $100r_g$ (where the gravitational radius is defined by
$r_g=GM/c^2$) or further from the black hole.  We note that MTR add
3\% systematic error to all data before comparing the various models.

In this {\it Letter}, we address the viability of the
absorption-dominated model by discussing a key constraint that is not
fully explored by MTR, i.e., the unavoidable connection between the
hard X-ray hump and fluorescent iron line production.  In both the
relativistic-reflection and the absorption-dominated models, the
low-energy side of the $\sim 20\keV$ hump is shaped by the
photoelectric absorption of iron.  Consequently, iron fluorescence
must accompany the 20\,keV hump; this is true even if there is no
optically-thick reflection occurring at all.  We argue that the only
way to make the presence of the hard X-ray hump compatible with the
extremely weak narrow iron line is to either strongly blur much of the
iron line (as in the relativistic-reflection model) or postulate a
very special geometry for the system.   

We present our arguments within the context of a re-analysis of the
January-2006 {\it Suzaku} observation of MCG--6-30-15.  Section~2
describes our reduction of these data.  In Section~3, we examine the
constraints imposed by relating the 20\,keV hump to the production of
iron line photons.  The meaning of these spectral constraints on the
geometry of the source are discussed in Section~4, and we draw our
conclusions in Section~5.  All error ranges are quoted at the 90\%
confidence level for one interesting parameter ($\Delta\chi^2=2.71$).

\section{The Suzaku data}

To illustrate the iron line constraints discussed in this paper in
concrete terms, we re-analyze the combined data of the three {\it
Suzaku} observations of MCG--6-30-15 taken 2006 January 9--14
(150\,ks), 2006 January 23--26 (99\,ks) and 2006 January 27--30
(97\,ks).  These data have been previously presented by Miniutti et
al. (2007). MCG--6-30-15 was placed at the nominal XIS aimpoint, and
all four X-ray Imaging Spectrometers (XIS0--3) were operational and
collecting data.  However, following Miniutti et al. (2007) we only
consider data from XIS2 and XIS3 which appear to show the fewest
calibration artifacts.  We also utilize data from the Hard X-ray
Detector (HXD) PIN.

Reduction started from the cleaned Version-2 data products, and data
were further reduced using FTOOLS version 6.4 according to the
standard procedure outlined in the ``Suzaku Data Reduction (ABC)
Guide''.  The standard filtering resulted in 338\,ksec of ``good''
data from each XIS.  Spectra were extracted from XIS2 and XIS3 using a
circular region of radius 4.3\,arcmin centered on MCG--6-30-15.
Background spectra were obtained from circular source free regions
(avoiding the calibration sources) around the chip edges.  Response
matrices and effective area curves were generated using the {\tt
xisrmfgen} and {\tt xissimarfgen} tools (called via the {\tt xisresp}
script provided to us by Keith Arnaud).  We use 100,000 photons per
energy bin during the construction of the effective area files.  We
also utilize HXD/PIN data in this paper.  Standard filtering resulted
in 274.2\,ksec of ``good'' {\it Suzaku}-HXD/PIN data from which a
spectrum was constructed.  A PIN background spectrum was produced that
models non-cosmic background.  The Cosmic X-ray Background was
included as an additional component to the spectral model as described
in the Suzaku Data Reduction guide.  Overall, our reduction repeats
that of Miniutti et al. (2007) except for the use of updated
instrumental response matrices and (importantly) HXD/PIN background
models.

To avoid complexities associated with the uncontroversial warm
absorption components (which possess significant opacity only below
2\,keV), we restrict all spectral modeling discussed here to energies
greater than 2.5\,keV.  Thus, we examine the 2.5--10\,keV XIS2/XIS3
spectra jointly with the 14--60\,keV HXD/PIN spectrum.   

\section{Connecting the iron line and the 20\,keV bump}

\begin{figure*}
\begin{center}
\hbox{
  \includegraphics[width=0.32\textwidth,angle=-90]{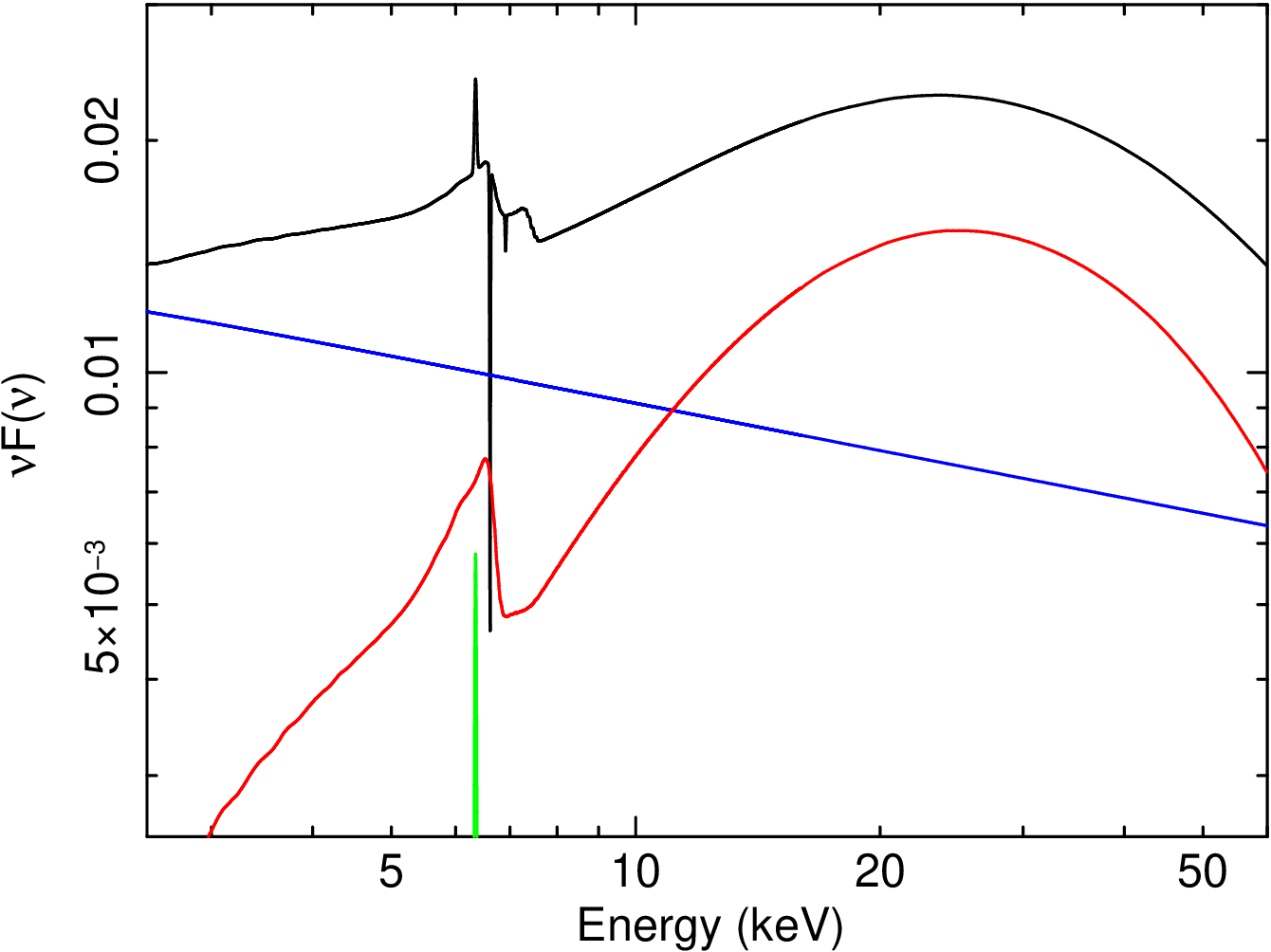}
  \hspace{0.5cm}
  \includegraphics[width=0.32\textwidth,angle=-90]{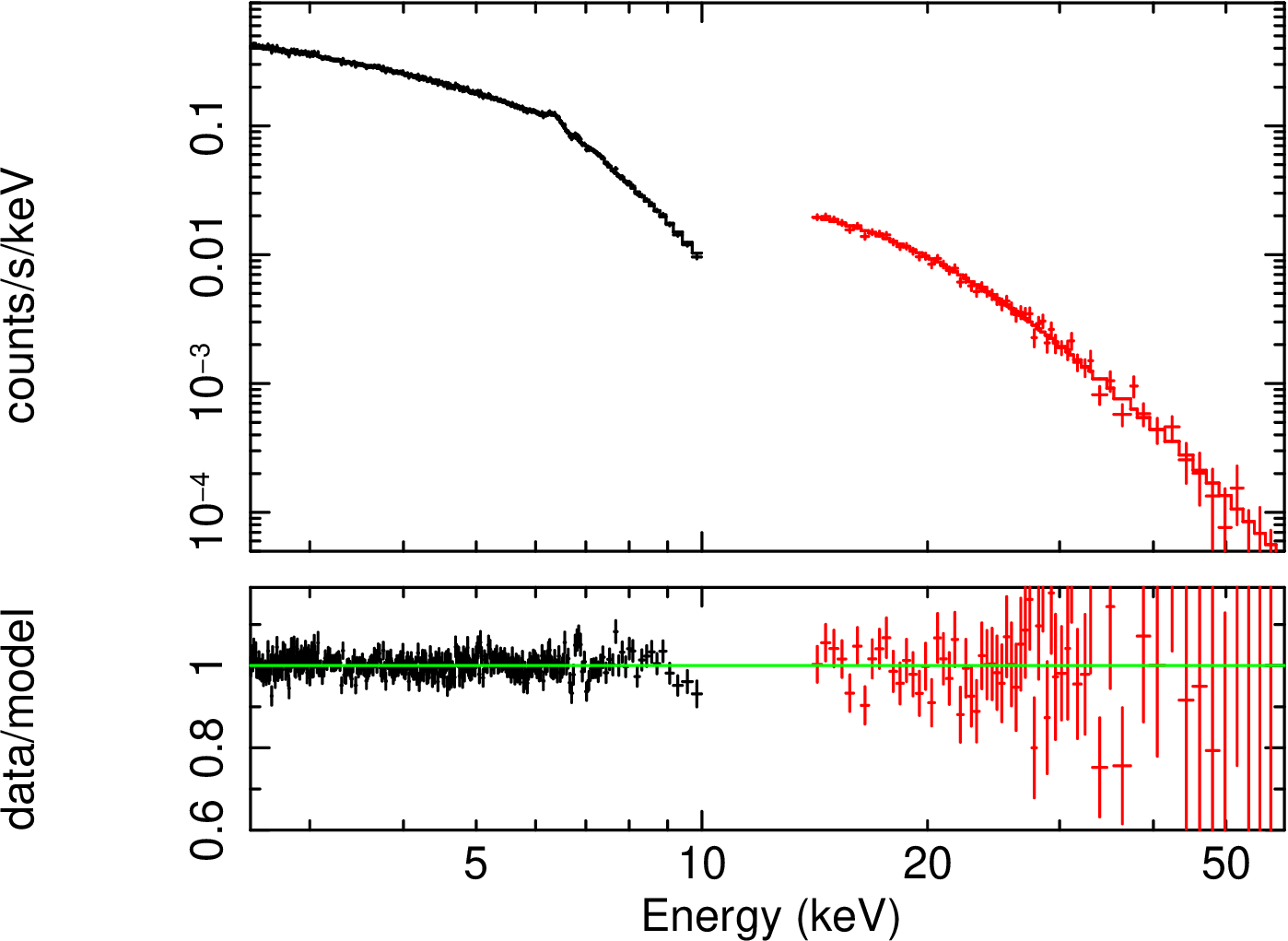}
}
\vspace{0.5cm}
\hbox{
  \includegraphics[width=0.32\textwidth,angle=-90]{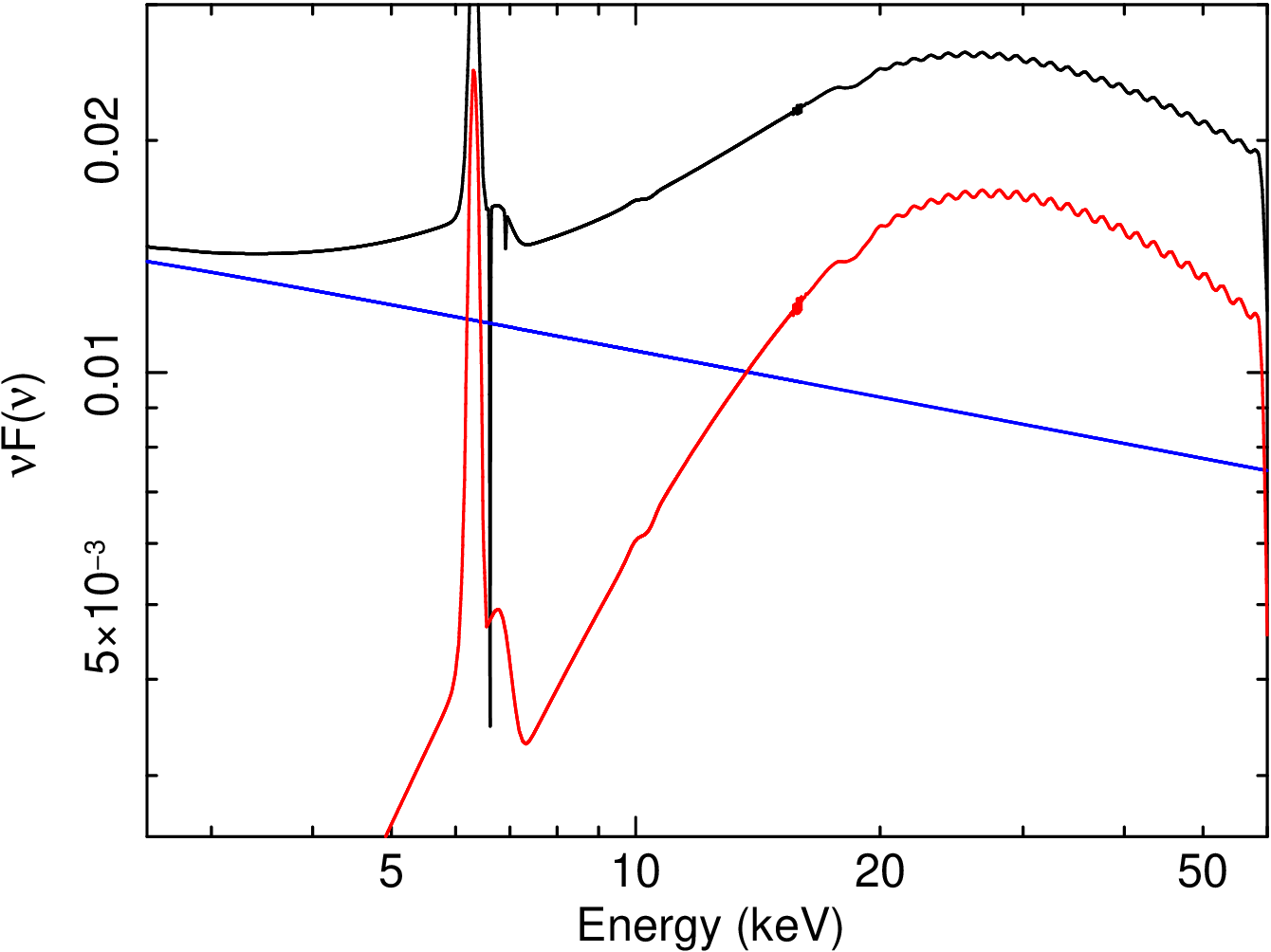}
  \hspace{0.5cm}
  \includegraphics[width=0.32\textwidth,angle=-90]{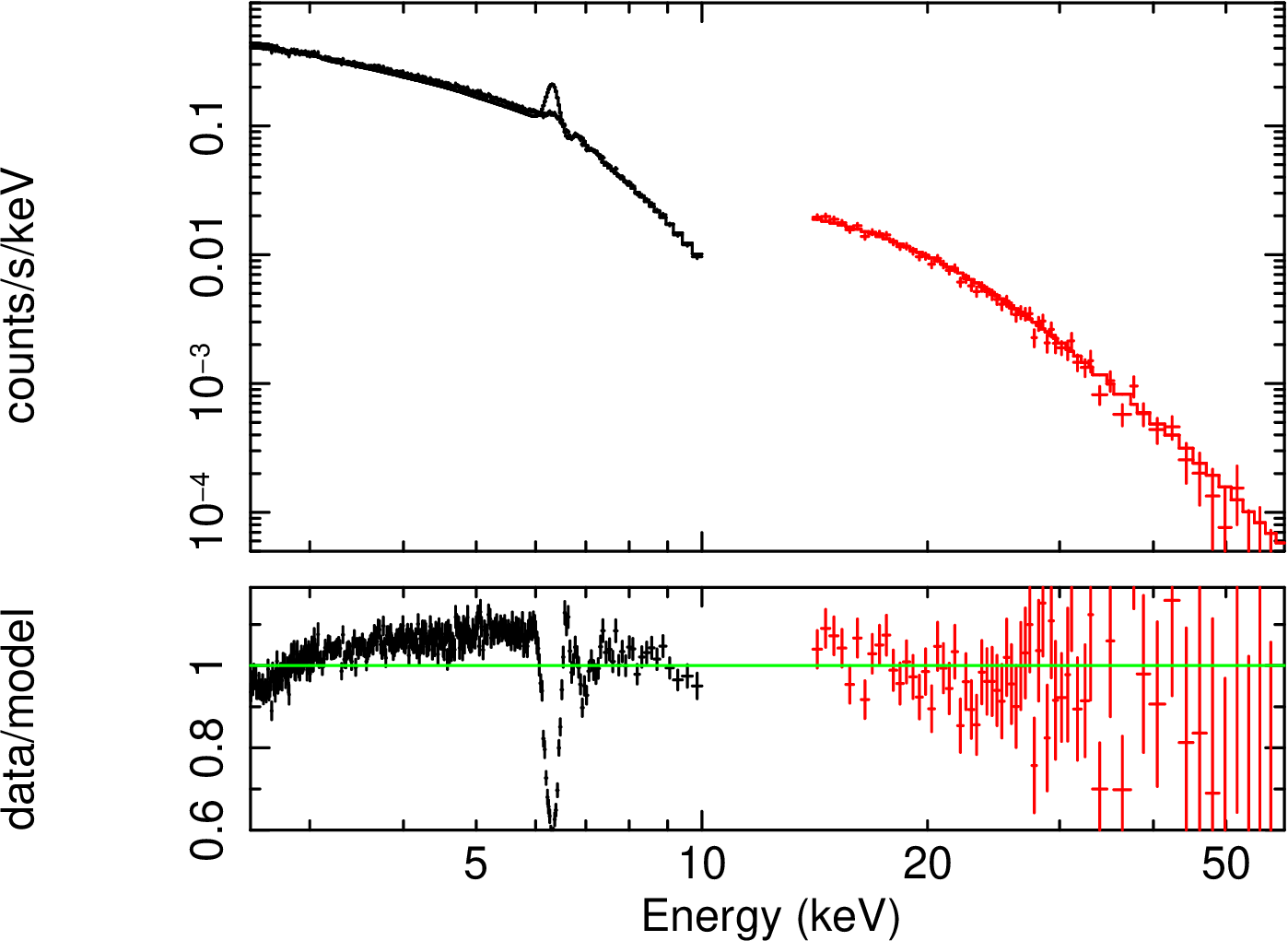}
}
\vspace{0.5cm}
\hbox{
  \includegraphics[width=0.32\textwidth,angle=-90]{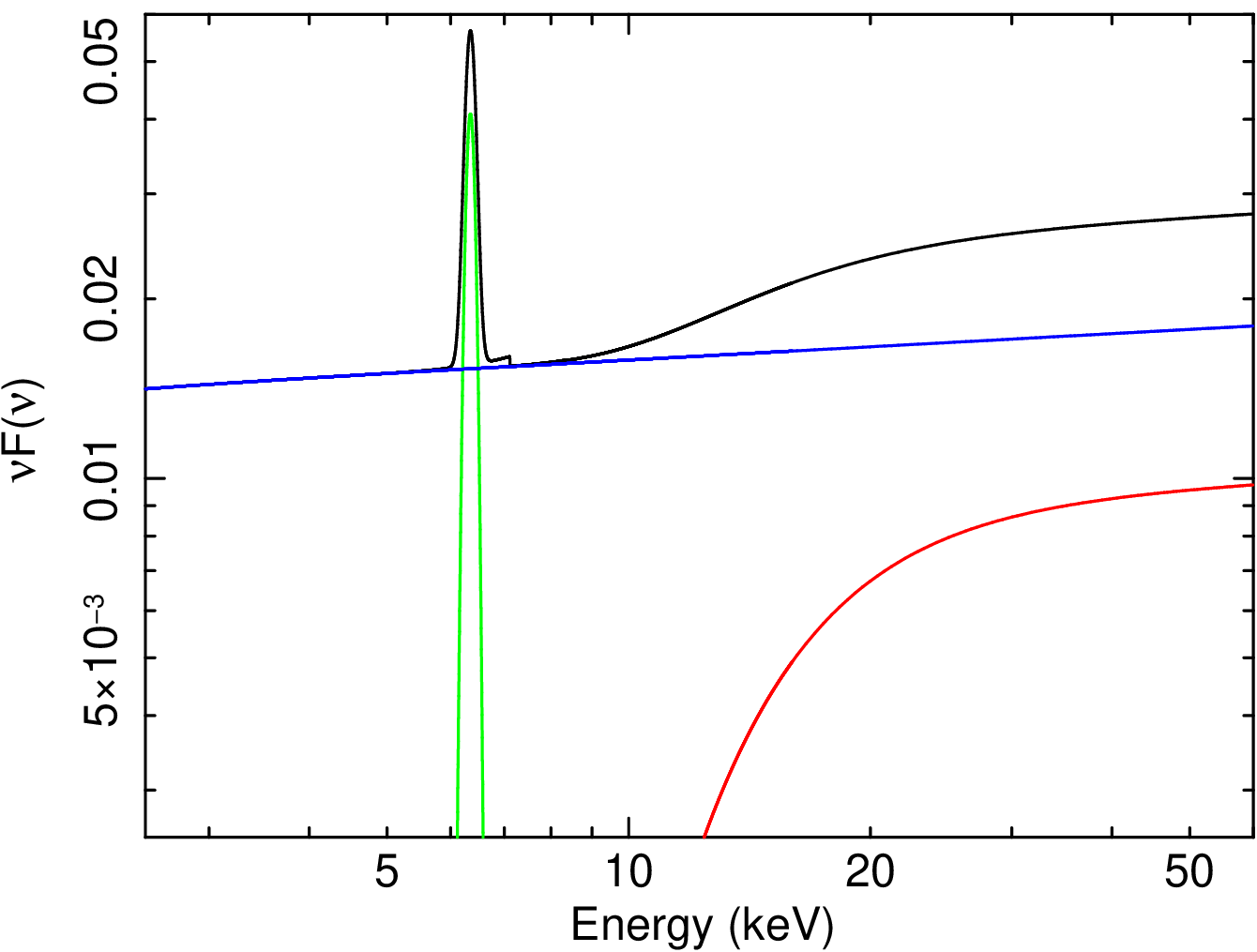}
  \hspace{0.5cm}
  \includegraphics[width=0.32\textwidth,angle=-90]{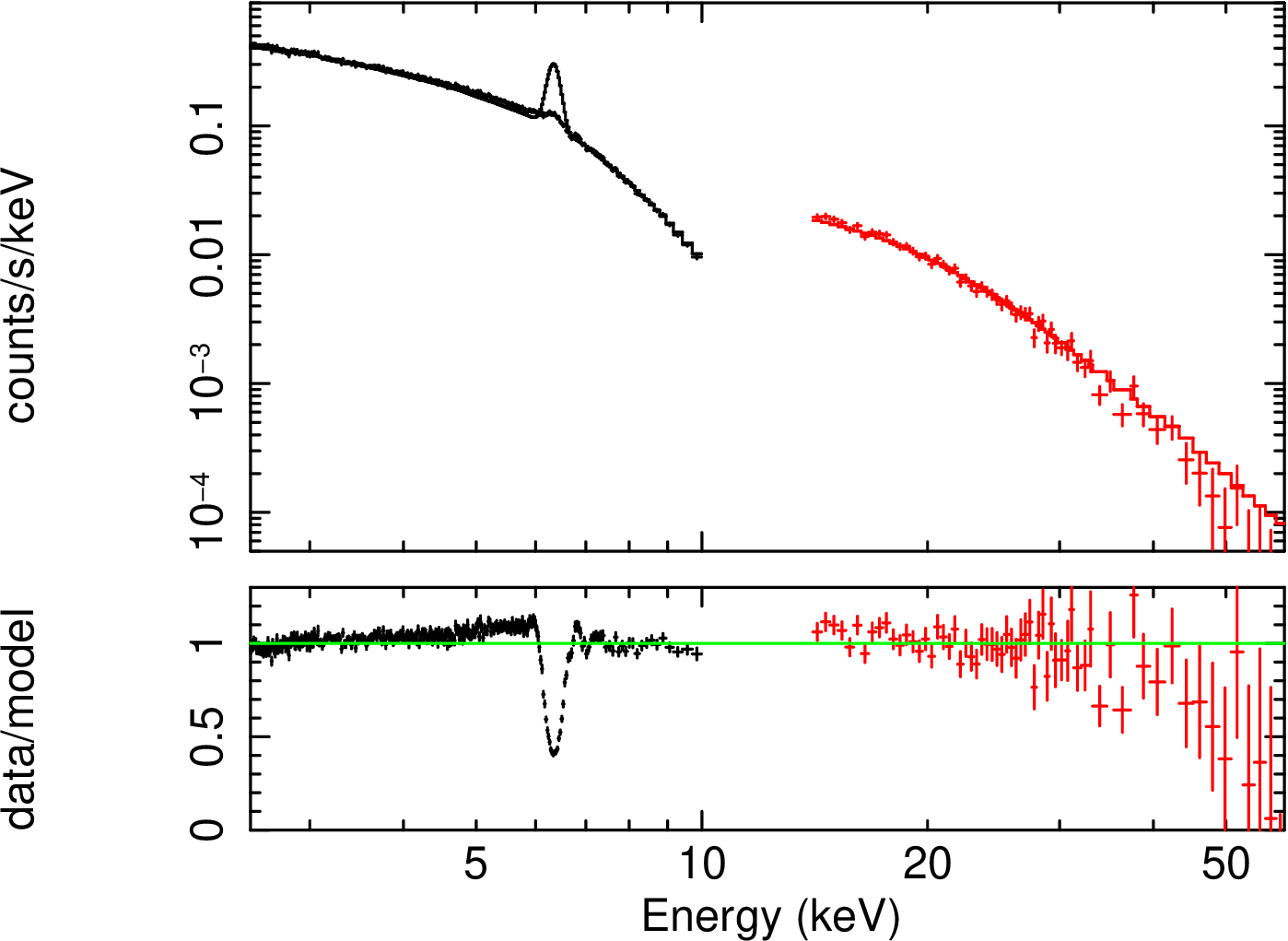}
}
\caption{{\it Top panels : }Relativistic reflection model (left)
together with a fit of this model to the 2.5--60\,keV {\it Suzaku}
data.  {\it Middle panels : }Reflection model in which the inner disk
has been fixed at $r_{\rm in}=200r_g$ (left) together with a fit of
this model to the 2.5--60\,keV {\it Suzaku} data.  {\it Bottom panels
: }Absorption-dominated model including the narrow iron line predicted
to accompany the photoelectric absorption (left) together with a fit
of this model to the 2.5--60\,keV {\it Suzaku} data.  See Section~3 of
the text for detailed descriptions of these models and the fits.}
\end{center}
\end{figure*}

Figure~1 (top panels) shows a fit of the relativistic-reflection model
to these {\it Suzaku} data.  In detail, the model consists of a
power-law continuum together with an ionized disk reflection spectrum
which has been blurred under the assumption that it originates from
the inner regions of an accretion disk around a near-maximally
rotating black hole.  We employ the ionized reflection models of Ross
\& Fabian (2005) and the relativistic smearing kernel of Laor (1991).
Note that Ballantyne et al. (2001) has demonstrated that diluted
constant-density ionized disk reflection models such as those
presented by Ross \& Fabian (2005) are a very good approximation to
the reflection expected from the surface layers of an accretion disk
in vertical hydrostatic equilibrium.  Following Brenneman \& Reynolds
(2006) and Miniutti et al. (2007), we also explicitly model
statistically significant but weak narrow emission/absorption features
using narrow Gaussians centered at 6.4\,keV (cold Fe fluorescent
emission), 6.67\,keV (He-like Fe absorption) and, 6.97\,keV (H-like Fe
absorption).  The Ross \& Fabian (2005) reflection models do not
include the K$\beta$ line of cold iron at 7.1\,keV; our spectral model
explicitly includes an additional 7.1\,keV line which is smeared
together with the ionized reflection model, and we see evidence for
such a line in the data.  The best-fit model is comparable to that
derived by Miniutti et al. (2007); the photon index of the primary
power-law is $\Gamma=2.09^{+0.04}_{-0.08}$, the reflector has iron
abundance $Z=1.8^{+0.3}_{-0.2}Z_\odot$ and ionization parameter
$\xi<35\erg\ps\cm$, and the relativistic smearing implies an
inclination of $i=31\pm 1^\circ$, an inner disk radius of $r=1.7\pm
0.1r_g$ and emissivity profile which breaks from $\beta=5.7\pm 0.7$ to
$\beta=2.74^{+0.15}_{-0.10}$ at $r_{\rm br}=3.8^{+0.6}_{-0.5}r_g$.

Of greatest importance to our current argument is the fact that a
powerful iron emission line accompanies the strong reflection hump.
As demonstrated in Fig.~1 (top panels), the presence of such a strong
line is compatible with the data only because it is strongly smeared
by the relativistic effects.  We demonstrate this more explicitly by
fitting the X-ray hump with a reflection model that does not include
any relativistic effects, as shown in Fig.~1 (middle panels).  More
precisely, starting with our relativistic-reflection model, we force
an inner disk radius of $r_{\rm in}=200r_g$ and refit the spectrum
ignoring the range of energies applicable to the broad iron line
(3.5--8\,keV).  As well as leaving a broad residual in the 3--7\,keV
band (corresponding to the broad iron line in the
relativistic-reflection model), this model vastly overpredicts the
narrow iron line.

Primarily motivated by this line-overproduction problem, MTR advocate
a model in which the hard X-ray hump is mainly shaped by direct
absorption rather than reflection.  However, as we shall now show, an
absorption-dominated model is still subject to constraints based on
iron line production.

To underscore our argument, we take an extreme form of the MTR model
that should be optimized to produce the least iron line emission; we
assume the complete absence of X-ray reflection and suppose that the
hard X-ray hump is shaped entirely by photoelectric absorption of the
primary X-ray emission by a cold, high column density absorber.  Of
course, the fact that we see plenty of emission at low-energies shows
that a significant fraction of the emission must either scatter around
this absorber or leak through holes in this absorber, i.e., this is a
partial-covering absorber.  Again, we fit this model to the {\it
  Suzaku} data ignoring the 3.5--8\,keV band.  This fit implies a
column density of $N_H=(2.0^{+0.36}_{-0.29})\times 10^{24}\pcmsq$ and
a covering fraction of $f=0.35\pm 0.03$.  Taking this best fit model
and setting $f=0$, we determine that the partial covering absorber is
removing $7.3\times 10^{-4}\,{\rm ph}\ps\pcmsq$ from the spectrum in
the 7.08--20\,keV (rest-frame) band; essentially all of these photons
are removed by iron K-shell photoelectric absorption (which has a
threshold energy of $7.08\keV$).  Assuming for now that the absorber
covers fraction $f=0.35$ of the sky isotropically as seen by the X-ray
source, the resulting iron line emission is simply given by the
product of the photoelectric absorption rate and the fluorescent
yield; using the fluorescent yield appropriate for cold iron (0.347),
we estimate an iron line photon flux of $2.54\times 10^{-4}\,{\rm
  photons}\ps\pcmsq$.  For illustration, Fig.1 (bottom panels) shows
this partial covering model including a 6.4\,keV iron line of this
strength and a width of $\sigma=0.1\keV$.

The allowed covering fraction of this marginally Compton-thick
absorber can be deduced by comparing the line photon flux predicted
above with the observed values/limits.  If we allow the normalization
of the narrow iron line to be free in the fit to the {\it Suzaku}
data, the observed excess at 6.4\,keV (which, in the
relativistic-reflection model corresponds to the blue-horn of the
broad iron line) has a photon flux of $2.50\times 10^{-5}\,{\rm
photon}\ps\pcmsq$; this implies that the Compton thick absorber covers
only $0.1f=0.035$ of the sky as seen by the X-ray source.  This is
robust to the addition or omission of a second partial covering
absorber to mimic the broad iron line.  If, instead, we use the
strength of the narrow 6.4\,keV iron emission line seen by the {\it
Chandra}/HETG (Young et al. 2005), the corresponding photon flux
($1.24\times 10^{-5}\,{\rm photon}\ps\pcmsq$) requires a covering
fraction of only $0.05f=0.0175$.  Of course, if there is any other
source of iron fluorescence (e.g., from reflection by very
Compton-thick matter), these covering fractions must be considered
upper limits. 

The detailed absorption-dominated model discussed by MTR is rather
more sophisticated than our ``extreme'' case in the sense that the
hard X-ray bump is modeled by an ionized reflector (that does not
display any relativistic effects and hence is inferred to be at large
distance from the black hole) which is absorbed by a photoionized
absorber with $N_H\sim 5\times 10^{23}\pcmsq$.  By construction, the
iron line from the absorbed reflection spectrum describes the narrow
weak 6.4\,keV line seen in the {\it Chandra}/HETG spectrum.  Using the
best-fitting parameters listed in MTR, we have confirmed that the
absorber removes the same number of 7.08\,keV--20\,keV photons from
the observed spectrum as calculated above for our extreme case, to
within a 10\% accuracy.  Since the fluorescence yield of iron
increases with ionization state, the expected fluorescence (of
intermediate charge states) should be even more prominent than
described above, leading to even more restrictive constraints on the
covering fraction of the absorber.  In principle, resonant Auger
destruction (e.g., Band et al. 1990; Ross et al. 1996) within both the
photoionized reflector and the absorber may reduce the iron line
strength somewhat, but significant reduction (more than a factor of
two) requires a fine-tuning of the ionization parameters and the
velocity structure.

The constraint discussed here is also robust to uncertainties in the
iron abundance of the absorber since we are essentially comparing the
iron edge to the iron line.  Repeating the exercise above (with our
extreme model) using an absorber with an iron abundance of $3Z_\odot$
and $0.3Z_\odot$ (all other abundances cosmic) yields predicted iron
line strengths that are within 10\% of the cosmic abundance case.

Note that the covering fraction constraints deduced here become
stronger upper limits if it is assumed that the Compton thick absorber
is in the form of clumps in a wind which is itself absorbing and
giving rise to fluorescent emission. 

\section{Discussion}

Despite the fact that it is not included in the commonly employed
spectral models, it is an elementary fact that fluorescent emission
accompanies photoelectric absorption.  As we have seen, exploiting
this connection provides a powerful new constraint on the
absorption-dominated model for MCG--6-30-15.  In particular, the
marginally Compton-thick absorber responsible for the hard X-ray hump
at 20\,keV cannot subtend more than 2--4 per cent of the sky as seen
from the X-ray source or else it will overproduce the fluorescent iron
line.

Thus, the absorption-dominated model requires a puzzling geometry.
The rapid fluctuations of the continuum source on timescales as short
as $\sim 100\s$ (Reynolds 2000; Vaughan, Fabian \& Nandra 2003)
suggest an X-ray source with an extent of only $\sim 2r_g$ (where we
are assuming a central black hole mass of $1\times 10^7\Msun$).
However, in the MTR model, all of the absorbing structures are at
$100r_g$ or further.  If these structures are to only
partially-obscure the central compact X-ray source, either the
observer must be in a special location so that the line of sight to the
absorber skims the edge of the absorbing structure, or the absorber
must be a ``mist'' of small clouds significantly smaller than the size
of the X-ray source (in which case there is a confinement problem).
So, within the context of the MTR model, the most reasonable
assumption is to place an additional continuum source at a larger
distance, comparable to that of the absorbing structure.

But the results of this paper provide a further constraint; that the
marginally Compton-thick absorber responsible for the hard X-ray hump
can subtend no more than 2--4\% of the sky as seen by the X-ray
continuum source.  If the absorber is approximately co-spatial with
the continuum source (as would be most natural given the
partial-covering constraint) it is very difficult to envisage a
geometry in which this holds true.  The geometric challenge is even
greater if one wishes to hypothesize that a large fraction of AGN have
similar absorption-dominated models (MTR; Sim et al. 2008).

By contrast, the relativistic-reflection model does not require
unusual geometries or previously unrecognized structures within the
central engine.  The difference spectrum (i.e., the difference between
the spectrum when the source is bright to when it is faint) is well
described by a power-law modified by only the well-studied warm
absorption components required by grating spectra; neither the broad
iron line nor the reflection hump show up in the difference spectrum
(Fabian et al. 2002; Miniutti et al. 2007).  The absorption-dominated
model can accommodate this spectral variability, but only via
unexplained correlations between the covering fraction of the
absorbers and the continuum luminosity.  By contrast, strong
gravitational light-bending is a {\it required and inevitable}
physical process within the relativistic-reflection model and
naturally explains the lack of response of the reflection features to
the continuum variability (Reynolds \& Begelman 1997; Matt et
al. 2000; Miniutti et al. 2003; Miniutti \& Fabian 2004).  Hand in
hand with the variability, the gravitational light-bending of
continuum X-rays towards the disk explains the fact that the strength
of the reflection hump is 2--3 times that expected from isotropic
illumination of a planar disk.  A prediction of the light-bending
model is that the broad iron line strength {\it is} correlated with
the continuum luminosity in low flux states; this was confirmed by
{\it XMM-Newton} spectroscopy of the June-2000 deep minimum state
(Reynolds et al. 2004).

Note that blurring a strong line produced by absorption, in say a
wind, by Compton downscattering also means that the high energy
continuum is also downscattered, to an unacceptable extent (Fabian et
al 1995). 

\section{Summary}

The relativistic-reflection model makes a bold claim, i.e., that we
are seeing measurable and quantifiable effects from matter within a
few gravitational radii of the black hole.  This gives us an
astrophysical tool of enormous power for probing black hole physics.
Thus, it is important to carefully assess alternative models such as
the absorption-dominated model of MTR.

{\it Suzaku} observations of the prototypical broad iron line AGN
MCG--6-30-15 clearly reveal a strong, hard X-ray hump (in the
10--30\,keV band) which must be due to either X-ray reflection or
absorption by a marginally Compton-thick absorber.  In either of these
cases, the low-energy side of the hump is shaped by the
K-shell photoelectric absorption of iron.  We consider the iron
fluorescence associated with the hard X-ray hump and come to the
following important conclusion; {\it the only way to make the presence
of the hard X-ray hump compatible with the extremely weak narrow iron
line is to either strongly blur much of the iron line (as in the
relativistic-reflection model) or postulate a very special geometry
for the system.}  This conclusion is robust to mild ionization of the
absorber as postulated in the MTR model (since the fluorescence yield
of iron and hence the predicted strength of the line increases with
ionization state).  Furthermore, since our argument amounts to
comparing the depth of the iron K-shall edge with the iron line, it is
robust to uncertainties in the iron abundance.  In the absence of a
specific geometry for the absorption-dominated model which satisfies
the constraints discussed here, the relativistic-reflection model is
favoured.

\section*{Acknowledgments}
CSR thanks the National Science Foundation for support under grant
AST06-07428.  ACF acknowledges the Royal Society for support. LWB
thanks ORAU for her current support at NASA's GSFC as a NASA
Postdoctoral Fellow.  GM thanks the Ministerio de Ciencia e
Innovaci\'on and CSIC for support through a Ram\'on y Cajal contract.
PU acknowledges funding from an STFC Advanced Fellowship.

\bibliographystyle{mnras}


\clearpage
\end{document}